\documentclass[aps,prl,twocolumn,showpacs,superscriptaddress]{revtex4-1}
\usepackage{amsmath,amssymb}
\usepackage{graphics,graphicx,color}
\usepackage{dcolumn,bm}
\usepackage{tikz}
\usepackage[colorlinks=true,citecolor=blue,urlcolor=blue]{hyperref}

\def\[{\left[}
\def\]{\right]}
\def\({\left(}
\def\){\right)}
\def\be{\begin{equation}}
\def\ee{\end{equation}}
\def\bea{\begin{eqnarray}}
\def\eea{\end{eqnarray}}

\newcommand{\gaug}
{\affiliation{Institute for Theoretical Physics, Georg-August-Universit\"at G\"ottingen, 37077 G\"ottingen, Germany}}

\begin{document}
\title{Multiple Types of Aging in Active Glass}

\author{Rituparno Mandal}%
\email[Email: ]{rituparno.mandal@uni-goettingen.de}
\gaug

\author{Peter Sollich}%
\email[Email: ]{peter.sollich@uni-goettingen.de}
\gaug
\affiliation{Department of Mathematics, King's College London, London WC2R 2LS, UK}

\begin{abstract}
Recent experiments and simulations have revealed glassy features in the cytoplasm, living tissues as well as dense assemblies of self propelled colloids. This leads to a fundamental question: how do these non-equilibrium (active) amorphous materials differ from conventional passive glasses, created either by lowering temperature or by increasing density? To address this we investigate the aging behaviour after a quench to an almost arrested state of a model active glass former, a Kob-Andersen glass in two dimensions. Each constituent particle is driven by a constant propulsion force whose direction diffuses over time. Using extensive molecular dynamics simulations we reveal rich aging behaviour of this dense active matter system: short persistence times of the active forcing lead to effective thermal aging; in the opposite limit we find a two-step aging process with {\em active athermal aging} at short times followed by {\em activity-driven aging} at late times. We develop a dedicated simulation method that gives access to this long-time scaling regime for highly persistent active forces.
\end{abstract}


\maketitle

Dense assemblies of particles show dynamical arrest when quenched to low temperatures or high density, with relaxation times becoming extremely long. Relaxation processes after a quench therefore show out-of-equilibrium behaviour, in particular a dependence on preparation history via the waiting time $t_{\rm w}$ since the quench. This phenomenon is known as {\textit{aging}}~\cite{berthier11}. It has been studied in great detail in spin glass models over the last few decades, both analytically and numerically~\cite{kisker96,cugliandolo94}. The first computational study on aging in structural glasses~\cite{barrat97} was performed on the {\textit{Kob-Andersen}} glass in three dimensions. When this system is quenched to a point close to dynamical arrest ($T=0.4$ in Lennard-Jones units), two-time overlap functions show a clear dependence on the waiting time $t_{\rm w}$, with typical relaxation times increasing as a power law of $t_{\rm w}$. While in the Kob-Andersen glass thermal effects are important, aging can occur also in athermal systems as discovered quite recently, with a distinct phenomenology~\cite{chacko19}.

\begin{figure}
\centering
\includegraphics[height = 0.9\linewidth]{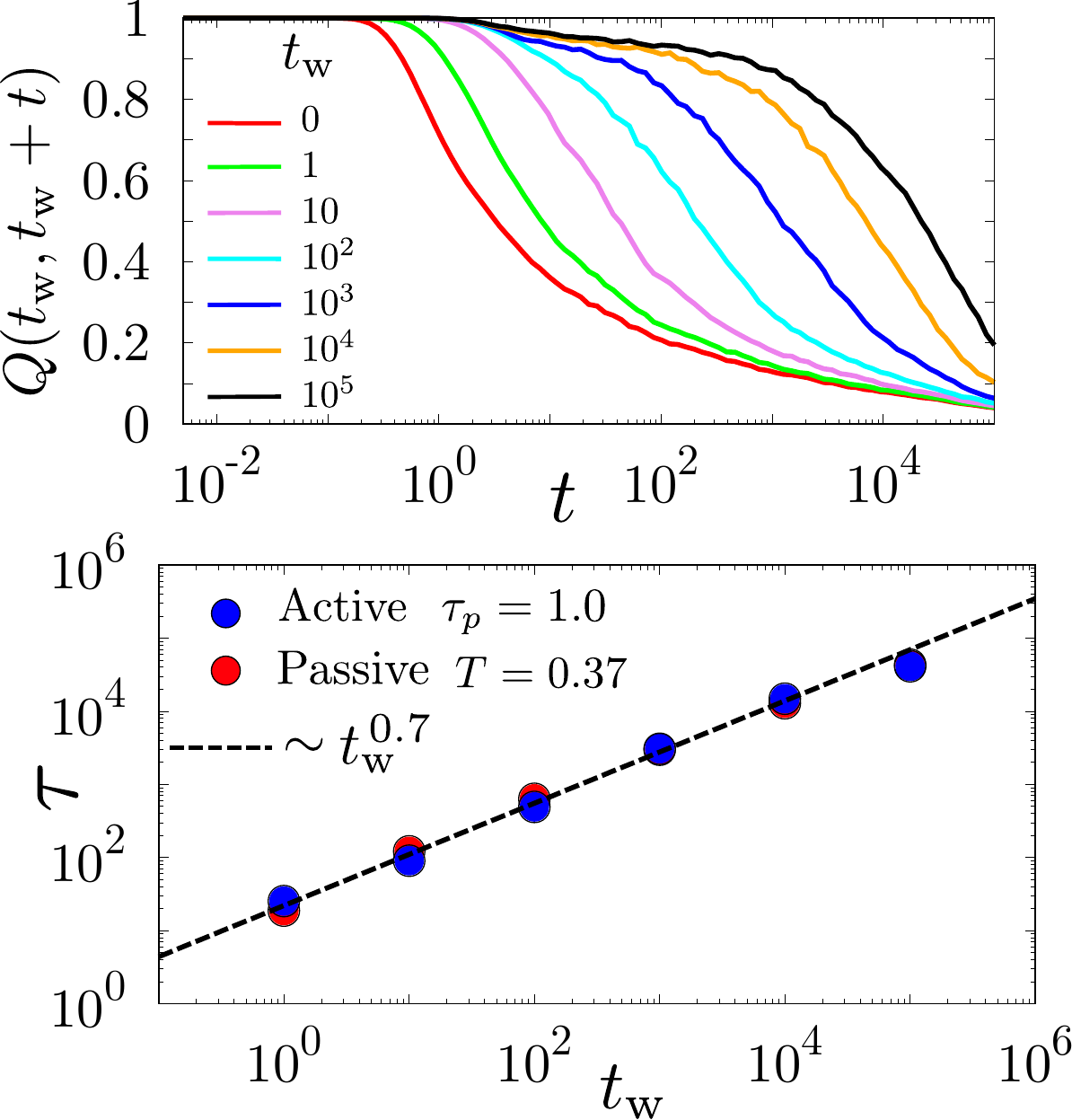}
\caption{(color online) (Top) Waiting time ($t_{\rm w}$) dependence of two-time overlap function $Q(t_{\rm w},t_{\rm w}+t)$ for $\tau_p=1$, following a quench in $f$ corresponding to an increase in steady state relaxation time from $\tau_\alpha=1$ to  $\tau_\alpha=10^7$ (extrapolated). (Bottom) The extracted age-dependent relaxation time $\tau$ versus $t_{\rm w}$ grows as a power law $\sim t^{0.7}_{\rm w}$ (dashed), in close agreement with the behaviour seen in a passive system after an equivalent temperature quench to $T=0.37$.}
\label{fig:smalltau}
\end{figure}

In the last decade the focus of many studies has shifted to understanding the behaviour of {\em self-propelled} particles at high densities, which form {\em active glasses}. Examples include dense cell assemblies in a tissue~\cite{angelini11}, extremely crowded cellular cytoplasm stirred by ATP-dependent activity~\cite{parry14} and self-propelled Janus colloids~\cite{leomach19}. Much work has also been done on the analytical~\cite{berthier13, szamel16, nandi18} and simulation~\cite{henkes11, ni13, berthier14, mandal16, bi16,mandal17} front; see Refs.~\cite{berthier19,janssen19} for reviews. 

The above recent results on active glasses raise a very fundamental question: how different is the dynamical behaviour of these glassy systems from that of conventional passive glasses? In this work we use aging, as a sensitive probe for slow glassy dynamics, to answer this question and differentiate dense active systems from their passive counterparts. To make this possible we design a close analogue of standard aging scenarios explored in experiments or numerical simulations on passive glasses, ``quenching'' an active supercooled liquid to a glassy phase by a sudden change of activity parameters. This allows us to isolate the underlying physics while avoiding additional effects from external perturbations such as oscillatory compression and expansion~\cite{janssen17}. Apart from shedding light on the key similarities and differences in the aging behaviour of active and passive glasses, our study can potentially motivate investigations of aging in recently discovered experimental active glasses~\cite{leomach19,bartolo19}, providing a deeper and more generic understanding of the off-equilibrium dynamics of dense active matter.

Our results below reveal unusually rich aging behaviour in a model active glass former. A key parameter is the persistence time $\tau_p$ of the active forces acting in the system~\cite{SI} for details, which along with the amplitude $f$ of the forces tunes the overall level of activity. For small $\tau_p$ we see aging behaviour similar to that in a thermal glass. This is due to the fact that rapidly fluctuating active forces act like Gaussian white noise in the limit $\tau_p \to 0$. In the opposite limit of large persistence time (formally $\tau_p\to\infty$) we observe a variant of athermal aging as recently described in passive systems~\cite{chacko19}, which we term ``Active Athermal Aging'' (AAA). This AAA has relaxation times growing as a power law of the age, with an aging exponent that decreases as the strength of active forcing is increased. For intermediate persistence time $\tau_p$, finally, we find a non-trivial two-step aging scenario. We explain this as a combination of AAA (for short timescales) and a new type of ``Activity Driven Aging'' or ADA (on long timescales well above $\tau_p$). We develop a computational approach to study such activity driven dynamics and demonstrate that this can also be deployed to understand the steady state behaviour (slow dynamics and intermittency) of glasses with highly persistent active forces~\cite{mandal19}.

We use molecular dynamics simulation to investigate off-equilibrium behaviour in an active glass, specifically a binary mixture of soft (Lennard-Jones) particles in two dimensions, with a number ratio of 65:35 to avoid crystallization. The passive limit of this model is known as the {\textit{Kob-Andersen}} glass~\cite{bruning09}. To make the system active, a self-propulsion force is added to the equation of motion of each particle. This has constant magnitude $f$ while its direction evolves diffusively, with a persistence time $\tau_p$ and independently for each particle. The addition of the self-propulsion force transforms the system into an active glass with rich dynamical properties~\cite{mandal19}. We follow Ref.~\cite{mandal19} in keeping the number density high ($\rho=1.2$) and focusing on the athermal limit ($T=0$). This leaves as the main tunable parameters the magnitude $f$ of the self-propulsion forces acting on each particle and the persistence time $\tau_p$ of this active forcing. 

\begin{figure}
\centering
\includegraphics[height = 0.67\linewidth]{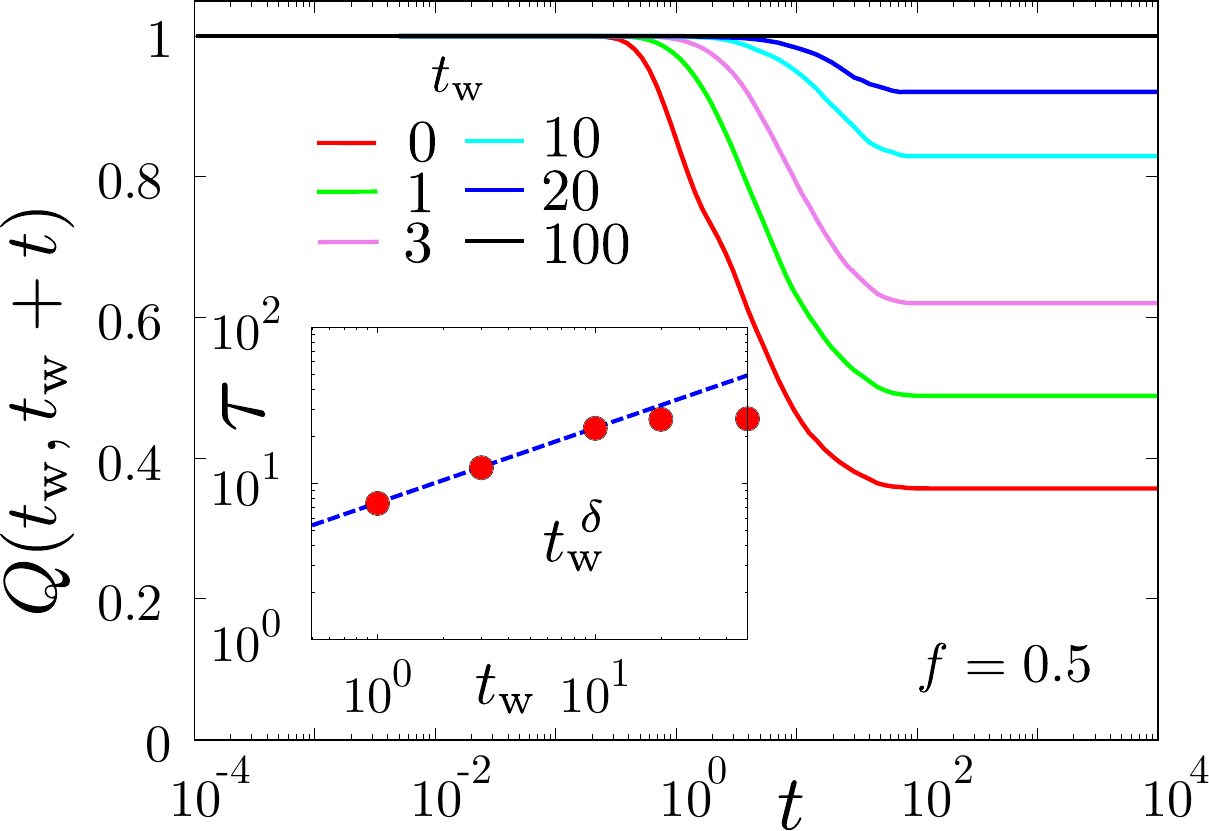}
\caption{(color online) Two-time overlap function $Q(t_{\rm w}, t_{\rm w}+t)$ as a function of $t$ for $\tau_p\to\infty$ and for different waiting times after a quench to $f=0.5$. Inset: Growth of relaxation time $\tau$ (extracted from the rescaled overlap function~\cite{SI}) with $t_{\rm w}$, fitted by a power law $\tau \sim t^{\delta}_{\rm w}$ with $\delta=0.48$.}
\label{fig:qt}
\end{figure}

The equation of motion of the particles and the phase diagram of the model have been discussed in detail in~\cite{mandal19} and are summarized in the SM~\cite{SI} for completeness. To observe aging we perform activity quenches at fixed persistence time $\tau_p$. To make results at different $\tau_p$ comparable, we proceed by measuring the relaxation time $\tau_\alpha$ as a function of $f$, in the liquid regime at large $f$. The initial value of $f$ is then chosen as the one where $\tau_\alpha=1$. To determine the post-quench value, we extrapolate the dependence $\tau_\alpha(f)$ into the glass regime and identify the $f$ for which $\tau_\alpha=10^7$, well beyond the longest timescales we can probe~\cite{SI}.

We start by looking at the small persistence time limit, taking specifically $\tau_p=1$. Following the protocol above we quench from a steady state at $f=5.67$ (where $\tau_\alpha=1$) to $f=1.38$ (with extrapolated $\tau_\alpha=10^7$) at time zero. We then wait for a time $t_{\rm w}$ and monitor the two-time overlap function $Q(t_{\rm w},t_{\rm w}+t)$. This measures the overlap between the particle configurations at the two times, specifically the fraction of particles that have moved less than some fixed distance~\cite{SI}. The results in Fig.~\ref{fig:smalltau} for different waiting times $t_{\rm w}$ show that the variation of $Q(t_{\rm w},t_{\rm w}+t)$ with age is similar in nature to the one observed in a thermal (passive) system. To quantify the aging we extracted for each $t_{\rm w}$ a relaxation timescale $\tau$ using the definition $Q(t_{\rm w}, t_{\rm w}+\tau)=1/e$. The waiting time dependence of the resulting relaxation time $\tau(t_{\rm w})$ is seen to be a power law with exponent $\delta_{\rm th} \sim 0.7$. This agrees qualitatively with observations made by Kob {\textit{et.\ al.}}~\cite{barrat97} in an analogous passive model in three dimensions. For a more quantitative comparison we performed aging simulations on our glass former in the passive limit ($f=0$) but at nonzero temperature. The initial condition was chosen as  equilibrium at high temperature $T=2.0$ (with again $\tau_\alpha=1$) and then the system was quenched to $T=0.37$ (a point equivalent to activity parameter $f=1.38$ at $\tau_p=1$~\cite{SI}). The resulting relaxation times as a function of waiting time $t_{\rm w}$, shown in Fig.~\ref{fig:smalltau}, essentially coincide with those of the active system with $\tau_p=1$. The shape of the overlap function decay also matches~\cite{SI}, clearly establishing that the active system has effectively thermal aging behaviour in the limit of small persistence time. The physical intuition that rationalizes this observation lies in the fact that rapidly fluctuating (small $\tau_p$) active forces are essentially equivalent to thermal white noise, in fact exactly so for $\tau_p\to 0$.

\begin{figure}
\centering
\includegraphics[height = 0.68\linewidth]{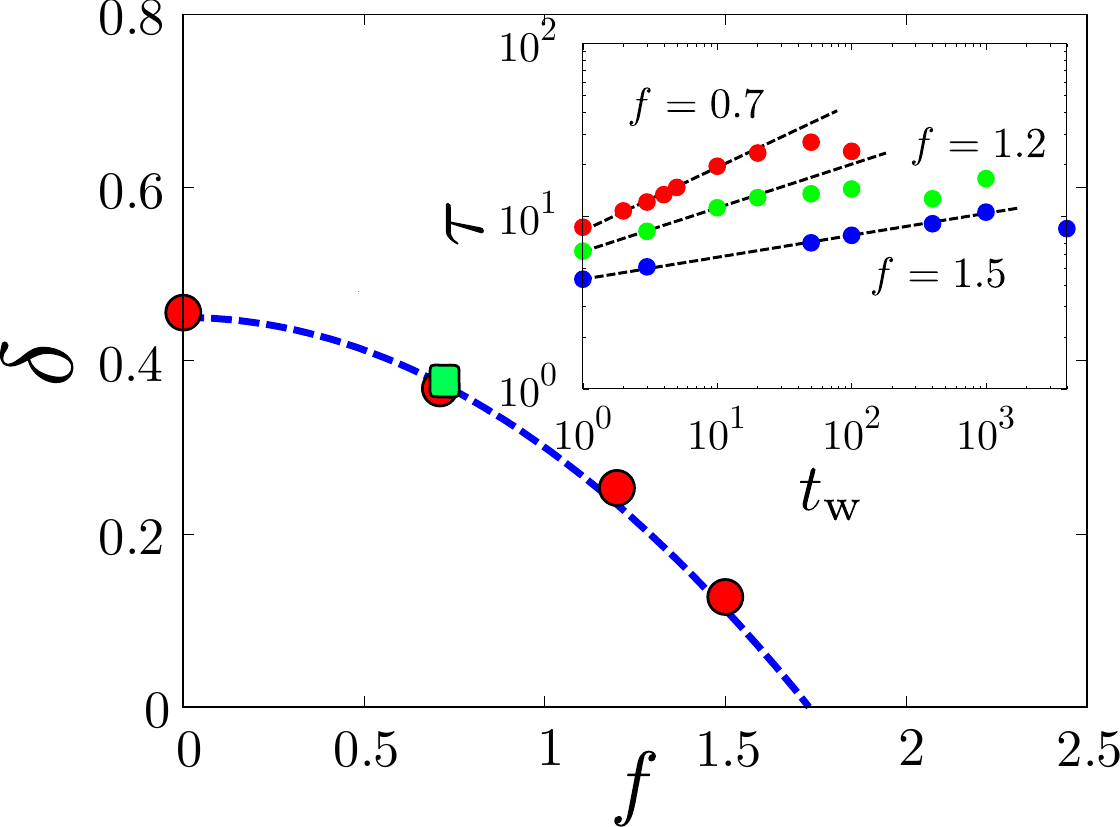}
\caption{(color online) Aging exponent $\delta$ for AAA ($\tau_p\to\infty$) for a range of post-quench values of active forcing $f$ up to  $f_c \approx 1.6$, exhibiting a roughly parabolic decrease (dashed blue line, guide to eye).
Green square: Aging exponent for quench to $f=0.72$ extracted from the short time dynamics at $\tau_p=10^4$. Inset: Growth of relaxation time $\tau$ as a function of $t_{\rm w}$ for several post-quench values of active forcing $f$.}
\label{fig:infinitetau}
\end{figure}

We next consider the opposite case of infinite persistence time $\tau_p$, quenching from $f=2.95$ where $\tau_\alpha=1$. As $\tau_\alpha(f)$ diverges discontinuously for $\tau_p\to\infty$ at some $f_c$ where the system becomes jammed (see ~\cite{SI, mandal19}), we cannot extrapolate $\tau_\alpha(f)$ smoothly and so quench to a range of different values below $f_c$. The resulting two-time overlap function $Q(t_{\rm w},t_{\rm w}+t)$ decays to a plateau whose height increases with increasing waiting time $t_{\rm w}$ (see Fig.~\ref{fig:qt}). One can extract a relaxation timescale for the decay to this plateau by scaling $Q(t_{\rm w}, t_{\rm w}+t)$ appropriately~\cite{SI}; this again exhibits a power law growth as a function of waiting time, $\tau \sim t^{\delta}_{\rm w}$. The aging exponent $\delta$ shows a smooth, approximately parabolic, decrease as the post-quench value of the active forcing $f$ varies from zero to $f_c$. Now, for a quench to $f=0$ the post-quench dynamics is that of a passive athermal system, so the active aging behaviour we observe for $\tau_p\to\infty$ and quenches to $f>0$ is a smooth continuation of (passive, $f=0$) athermal aging as studied very recently~\cite{chacko19}. We therefore term it \textit{Active Athermal Aging} (AAA). Further support for the connection to passive athermal aging comes from the fact that for the kinetic energy as studied in~\cite{chacko19}, we find a power law decay in time~\cite{SI} just as in the passive case~\cite{chacko19}. The corresponding exponent decreases with the post-quench value of $f$ in a manner analogous to Fig.~\ref{fig:infinitetau}, although the precise relation between these exponents remains an open question for future work. For values of $f$ close to $f_c$ the appearance of a plateau in the kinetic energy decay hints at different physics that also remains to be explored, in possible analogy with transient flow occurring in the creep of yield stress fluids held at stresses below yield~\cite{barrat18,liu18}. 

\begin{figure}
\centering
\includegraphics[height = 0.68\linewidth]{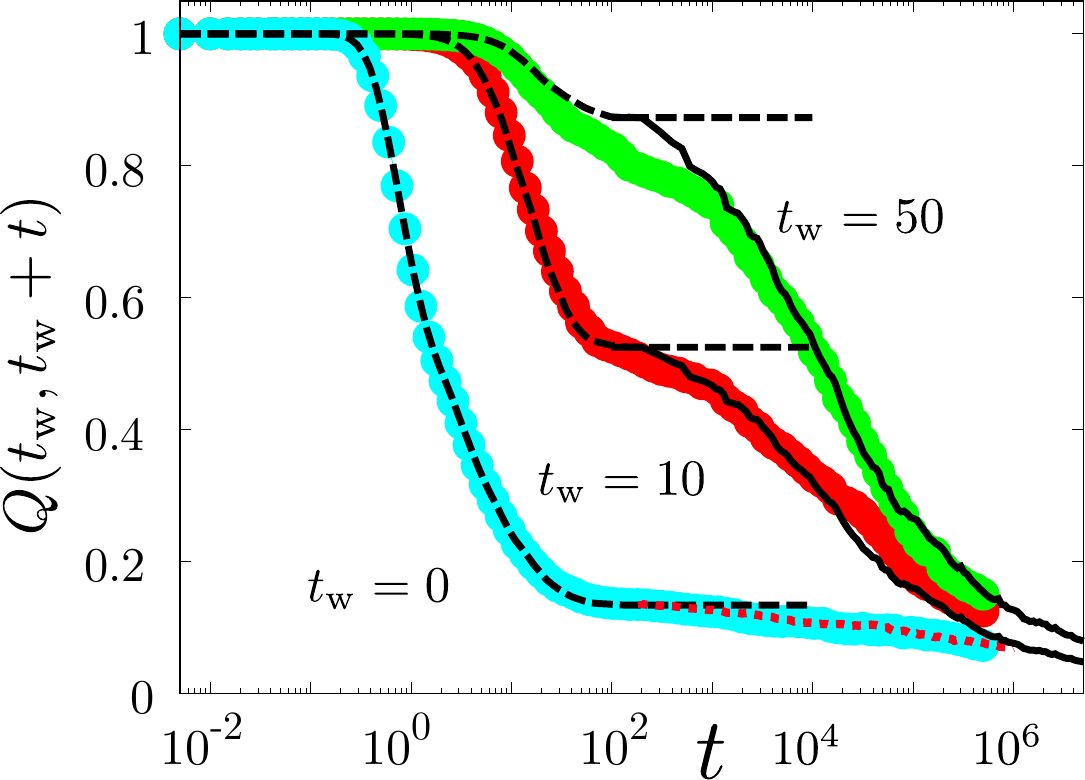}
\caption{(color online) Two-time overlap function  $Q(t_{\rm w},t_{\rm w}+t)$ (coloured points) for intermediate persistence time  $\tau_p=10^4$, after a quench to $f=0.72$. Black dashed lines: AAA dynamics for the same quench and initial conditions but with $\tau_p\to \infty$. Black solid lines: Prediction from ADA for $t^{\prime}_{\rm w} \ll 1$, scaled by the plateau reached during AAA. Brown dashed line: ADA for $t^{\prime}_{\rm w}=0$ (without scaling).}
\label{fig:largetau}
\end{figure}

For intermediate persistence time we follow the general quench protocol described above,
quenching to a value of $f$ with extrapolated steady state relaxation time of $\tau_\alpha=10^7$. We observe in Fig.~\ref{fig:largetau} a non-trivial two step decay of the two-time overlap function $Q(t_{\rm w},t_{\rm w}+t)$. To understand this we note that for $t\ll \tau_p$ the active forces remain essentially unchanged, so the dynamics should be close to that for $\tau_p\to\infty$, {\em i.e.}\ of AAA type. We confirm this in Fig.~\ref{fig:largetau} by explicit comparison with the results for infinite persistence time, which closely reproduce the initial decay of $Q(t_{\rm w},t_{\rm w}+t)$. This can be made more quantitative by again extracting a $t_{\rm w}$-dependent relaxation time for this initial decay of the two-time overlap function; we find that its growth is described by an exponent that is indistinguishable from the one for $\tau_p=\infty$ as long as $\tau_{\rm p}$ is large enough, as shown by the green square in Fig.~\ref{fig:infinitetau}.

On longer timescales $t \geq \tau_p$ we observe in Fig.~\ref{fig:largetau} a decay of $Q(t_{\rm w},t_{\rm w}+t)$ from the plateau. This clearly has no analogue in AAA and -- given the large $\tau_p=10^4$ -- also cannot be thermal aging. To access this distinct type of aging dynamics in conventional simulations is difficult as it requires looking at $t$ (and possibly $t_{\rm w}$) of the order of or larger than $\tau_p$. We therefore develop a method that directly allows us to see the behaviour on timescales of order $\tau_p$ for large $\tau_p$; formally it applies in the limit $\tau_p\to\infty$ taken at finite {\em scaled} times $t'=t/\tau_p$ and $t_{\rm w}' = t_{\rm w}/\tau_p$. In this way we can access a time regime that is practically impossible to reach in vanilla simulations, especially for very large $\tau_p$. The insight behind the method is the following: for $\tau_p\to\infty$, the relaxation of the particle configuration to a stable force-free state, for the given self-propulsion forces, takes negligible time relative to $\tau_p$. The particle coordinates therefore follow the changes in the active forces {\em adiabatically}. Accordingly we call this regime activity-driven dynamics, and refer to the aging that results as ADA. For the numerical implementation we repeatedly make small random changes in the orientation of the self-propulsion forces of the particles and then let the system find a new force-free state. Because for small displacements the active forces can be thought of as gradients of additional linear potential terms, this state is also a local minimum of the appropriately tilted energy surface. Each such timestep advances the {\em scaled} time $t'=t/\tau_p$ by some small amount, because it is this scaled time that the changes in self-propulsion direction depend on. In spirit this dynamics is similar to Athermal Quasistatic Shear~\cite{takeuchi78, takeuchi80, takeuchi81, lemaitre04, lemaitre06}, where the limit of small shear rate $\dot\gamma$ is taken by allowing the particle configuration to relax fully after each small increment of the strain $\gamma=\dot\gamma t$.

\begin{figure}
\centering
\includegraphics[height = 0.41\linewidth]{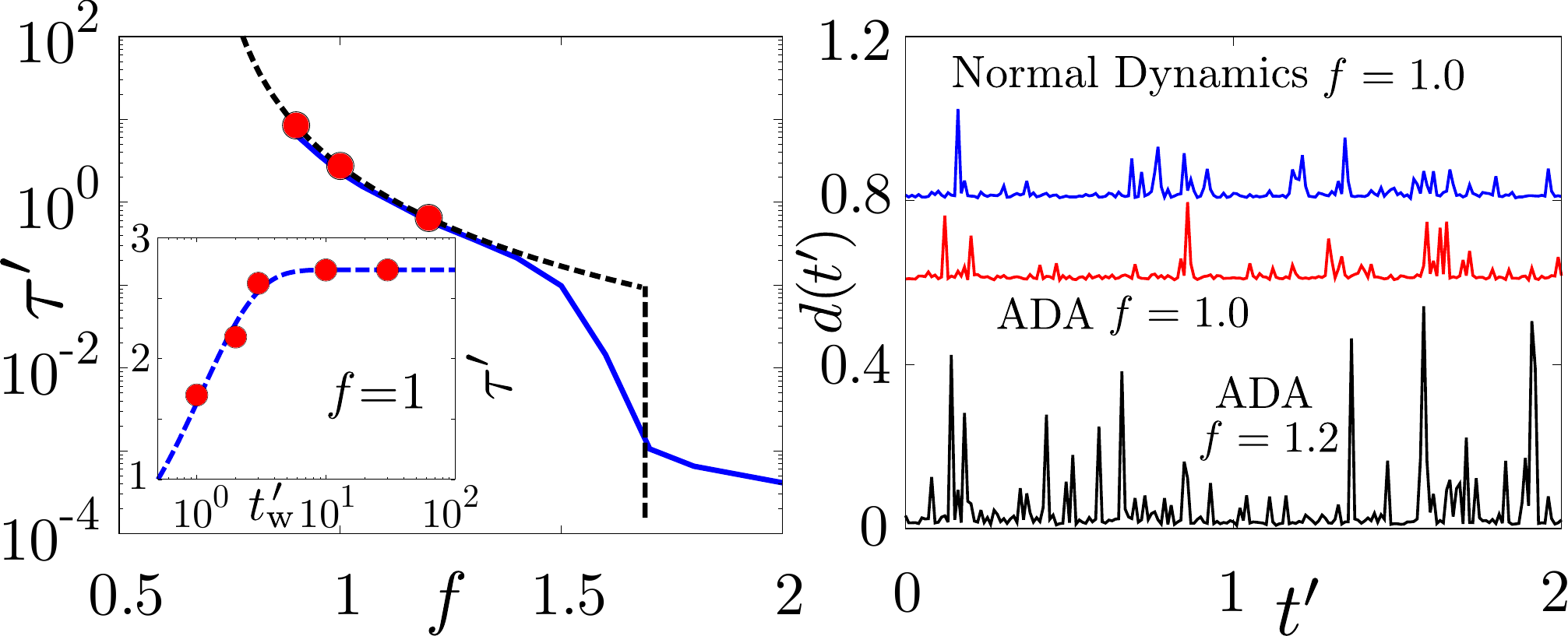}
\caption{(color online) Left: Scaled steady state relaxation time for activity driven dynamics (red points) and usual dynamics with $\tau_p=10^4$ (blue line) for different $f$. The vertical black dashed line indicates $f_c$, the point at which (for large $\tau_p$) the relaxation timescales change from scaling linearly with $\tau_p$ to being independent of $\tau_p$~\cite{mandal19}.
Inset: Scaled relaxation time for activity driven dynamics after a quench to $f=1.0$ as a function of scaled waiting time $t^{\prime}_{\rm w}$, showing initial aging before the (here reasonably short) steady state relaxation time is reached. Right: Time series of root mean-squared  displacements $d(t')$ for usual (blue) and activity driven dynamics (red) in steady state at $f=1.0$ show qualitatively similar behaviour, {\textit{e.g.}}\ intermittent dynamics with clearly visible bursts~\cite{mandal19}. Similar data for $f=1.2$, generated using activity driven dynamics, shows an increase in both the strength and frequency of bursts.}
\label{fig:ADD}
\end{figure}

For validation of the activity-driven dynamics approach we first ensure that results have converged towards the two required limits of infinitesimal steps of scaled time $t'$ and zero threshold for residual forces after the minimization~\cite{SI}. We then extract steady state quantities from the activity-driven dynamics at large $f$ 
and compare to conventional simulations with large but finite persistence time $\tau_p$. Fig.~\ref{fig:ADD} shows that both the relaxation time $\tau_\alpha$ and the intermittent character of particle displacements~\cite{mandal19} match well between the two approaches, establishing the effectiveness and accuracy of the method.

\begin{figure}
\centering
\includegraphics[height = 0.7\linewidth]{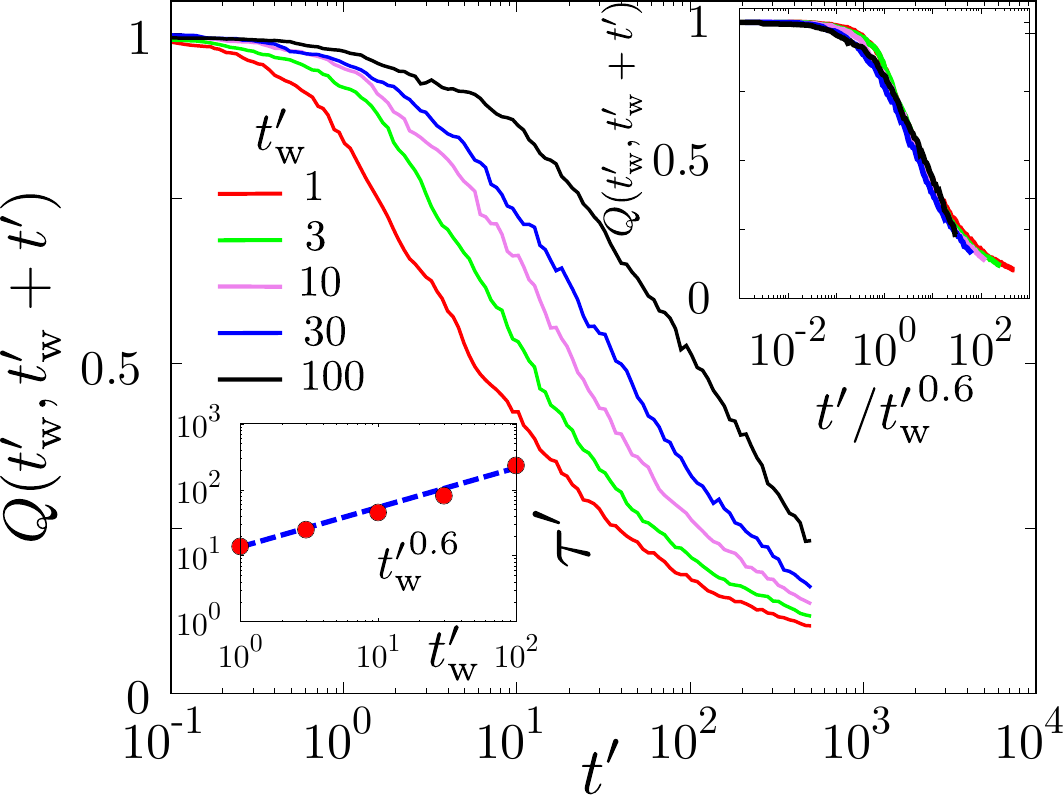}
\caption{(color online) Two-time overlap function $Q(t^{\prime}_{\rm w},t^{\prime}_{\rm w}+t^{\prime})$ from ADA, plotted against time difference $t^{\prime}$ for different scaled waiting times $t^{\prime}_{\rm w}$. Top inset: Data can be collapsed by an age-dependent scaling of $t'$. Bottom inset: Growth of the corresponding relaxation time $\tau^{\prime}$ as a power law in $t'_{\rm w}$. All times shown are in units of $\tau_p$ ($t'=t/\tau_p$ etc).}
\label{fig:ADA}
\end{figure}

Moving on to aging, we look at the two-time overlap function after a quench to a point where the extrapolated {\em scaled} relaxation time $\tau_\alpha^\prime=\tau_\alpha/\tau_p=10^3$; Fig.~\ref{fig:ADA} shows the resulting $Q(t^{\prime}_{\rm w},t^{\prime}_{\rm w}+t^{\prime})$. From this a scaled, $t_{\rm w}'$-dependent relaxation time can be extracted as before. This shows power law growth with an aging exponent $\delta^{\prime}\sim 0.6$ that is physically unrelated to -- and numerically different from -- the one for thermal aging. 
We note that this aging behaviour arises solely from ``internal'' activity in the form of self-propulsion at the particle scale; it is thus fundamentally different from aging or rejuvenation under external ({\em e.g.}\ periodic~\cite{janssen17}) perturbation. The self-propulsion effectively stirs the system and behaves locally as micro-shear, with ensuing plastic rearrangements and Eshelby stress propagation~\cite{mandal19}. Surprisingly, this kind of driving does {\em not} stop aging but in fact {\em facilitates} it. This is in stark contrast to driving by steady external shear, which destroys or ``interrupts'' any aging process as argued theoretically and seen in experiments and simulations~\cite{kurchan01, berthier01, abou03}. 

We finally show that ADA, which is derived for $\tau_p\to\infty$, rationalizes the late-time decay of $Q(t_{\rm w},t_{\rm w}+t)$ for finite $\tau_p$. For the case $t_{\rm w}=t_{\rm w}'=0$ we start ADA runs from the same initial condition as used for $\tau_p=10^4$; after scaling back to $t=t'\tau_p$ the predictions (brown dashed line in Fig.~\ref{fig:largetau}) match those from the conventional simulations very well.

Note that ADA by design does not predict the short-time decay to the plateau, which in the $\tau_p\to\infty$ limit gets compressed into a discontinuous drop in $Q$ from $Q=1$ to the plateau at $t'=0$. For $t_{\rm w}'>0$, on the other hand, ADA has already reached a force-free configuration at $t_{\rm w}'$ and the overlap $Q(t^{\prime}_{\rm w},t^{\prime}_{\rm w}+t^{\prime})$ with this configuration decays smoothly from $Q=1$ without a plateau (see Fig.~\ref{fig:ADA}). We therefore multiply the ADA predictions with the plateau predicated from the short-time (AAA) dynamics; this again gives an accurate account of the finite $\tau_p$-data (see Fig.~\ref{fig:largetau}). Overall this establishes the fact that the non-trivial two-step aging we see for large persistence times arises from a combination of AAA (on short timescales $t < \tau_p$) and ADA for longer times. 

In conclusion, we report unexpectedly rich aging behaviour in a model active glass. In the limit of small persistence time $\tau_p$ we see thermal-like aging that we can map quantitatively onto a passive glassy system quenched to low $T$. In the opposite limit of infinite persistence time we find  AAA, with an aging exponent decreasing as the post-quench amplitude $f$ of the active forcing increases and connecting smoothly to passive athermal aging~\cite{chacko19}. For intermediate persistence time we observe a two-step aging process scenario that can be understood as a combination of AAA and, on timescales of order $\tau_p$ and beyond, ADA. Remarkably, in the latter regime the stirring by self-propulsion forces does not destroy aging but in fact drives it. To gain access to this late-time regime we designed a custom simulation approach that will be useful also for broader investigations into the dynamics of systems with strongly persistent activity. Further work is being pursued to understand the aging exponents which, with values below unity, generally indicate sub-aging. Other interesting directions will be to study how intermittency in the steady state dynamics of the model~\cite{mandal19} manifests itself in the aging behaviour, and how our results relate to aging in lower density active gels reported recently~\cite{merrigan19}. It will also be worthwhile to explore the spatial signatures of the different types of aging and establish possible connections with two- and four-point density correlation functions. Our results should thus provide a useful foundation for further research into the out of equilibrium dynamics of dense active or living glassy systems.

We are grateful to Suzanne Fielding and J\"org Rottler for insightful discussions.

\end{document}


\title{Supplementary material for: Multiple Types of Aging in Active Glass}

\author{Rituparno Mandal}%
\email[Email: ]{rituparno.mandal@uni-goettingen.de}
\gaug

\author{Peter Sollich}%
\email[Email: ]{peter.sollich@uni-goettingen.de}
\gaug
\affiliation{Department of Mathematics, King's College London, London WC2R 2LS, UK}

\maketitle

\newpage

\subsection{Model and Phase Diagram}

We study a two-dimensional glass former made of soft Lennard Jones particles. The number ratio of particles ($A:B=65:35$) in the binary mixture, density $\rho=1.2$ and the interaction parameters ($\epsilon_{ij}$ and $\sigma_{ij}$) are chosen in such a way that without the self propulsion force the system behaves as a passive two-dimensional Kob-Andersen glass ~\cite{bruning09}. To include activity we add an active forcing term in the equation of motion for each particle,
\begin{equation}
m{\dot{\mathbf{v}}}_i=-\gamma \mathbf{v}_i +\sum_{j=1, j\neq i}^{N} \mathbf{f}_{ij} + f \mathbf{n}_i
\label{eq_of_motion}
\end{equation}
where $m$ is the mass, $\mathbf{v}_i$ is the velocity of the $i$-th particle and $\gamma$ is the friction coefficient. In the current work we keep $m=1$ and $\gamma=1$ throughout. In Eq.~(\ref{eq_of_motion}), $\mathbf{f}_{ij}$ is the Lennard-Jones interaction force acting between particle $i$ and $j$, 
$\mathbf{n}_i$ is the direction unit vector associated with the self-propulsion force of particle $i$ and $f$ is the magnitude of this force. The direction of propulsion, $\mathbf{n}_i$ follows a rotational diffusion equation with rotational diffusion constant $D_R \propto \tau_p^{-1}$ such that $\tau_p$ is the persistence time of the active forcing. The number of particles used in the simulation is $N=1000$. All data reported here have also been averaged over $100-1000$ independent realisations unless specified otherwise.

\begin{figure}
\centering
\includegraphics[height = 0.62\linewidth]{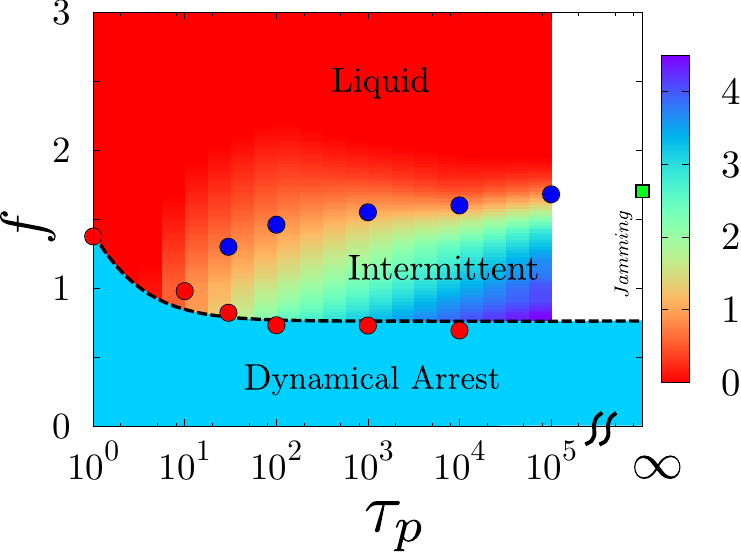}
\caption{(color online) Phase diagram of the model active glass (taken from Ref.~\cite{mandal19}). For small persistence time system undergoes liquid to glass transition and at infinite persistence time liquid to a jammed state as the active forcing $f$ is reduced. For intermediate and large persistence time ($\tau_p \sim 10^4$), for moderate values of active forcing $0.7<f<1.6$ the system shows an intermittent dynamical phase.}
\label{fig:phased}
\end{figure}

The model studied here~\cite{mandal19} has a rich phase diagram (see Fig.~\ref{fig:phased}) governed by the two key parameters $f$, the magnitude of the self-propulsion forces, and the persistence or ``activity decorrelation'' time $\tau_p$. For small $\tau_p$ the system undergoes a liquid to glass transition as $f$ is lowered. At infinite persistence time the system again behaves like a liquid at large $f$ but gets dynamically jammed around $f_c\approx 1.6$. For reasonably large but finite $\tau_p$, there exists an intermittent dynamical phase between the active liquid and the dynamically arrested phase. To understand the dynamical response of the system we define a two-time overlap function as (see e.g.~\cite{franz00}; the earlier aging study~\cite{barrat97} used a similar function of dynamical structure factor form):
\begin{equation}
Q(t_{\rm w}, t_{\rm w}+t) =\langle \frac{1}{N} \sum_{i} q(\mid{\bf r}_i(t_{\rm w}+t)- {\bf r}_i(t_{\rm w})\mid)\rangle
\end{equation}
where
\begin{equation}
q(x)=
\left\{
        \begin{array}{ll}
                1  & \mbox{if } x \leq a\\
                0  & \mbox{otherwise}
        \end{array}
\right.
\end{equation}
and $t_{\rm w}$ is the waiting time before the measurement starts. The sum in this definition is over particles and the angular brackets $\langle \rangle$ indicate averaging over independent initial realisations and the stochasticity in the dynamics. Note that no averaging over the time origin is involved as our focus is precisely on how the two-time overlap function evolves with waiting time $t_{\rm w}$. The parameter $a$, which sets the distance beyond which particles are counted as having moved, is chosen throughout as $a=0.3$ in units of $\sigma_{AA}$ where the interaction ranges of the different pairs of particle types are $\sigma_{AB}=0.8 \sigma_{AA}$ and $\sigma_{BB}=0.88 \sigma_{AA}$. 

\subsection{Aging Protocol} 

\begin{figure}
\centering
\includegraphics[height = 0.62\linewidth]{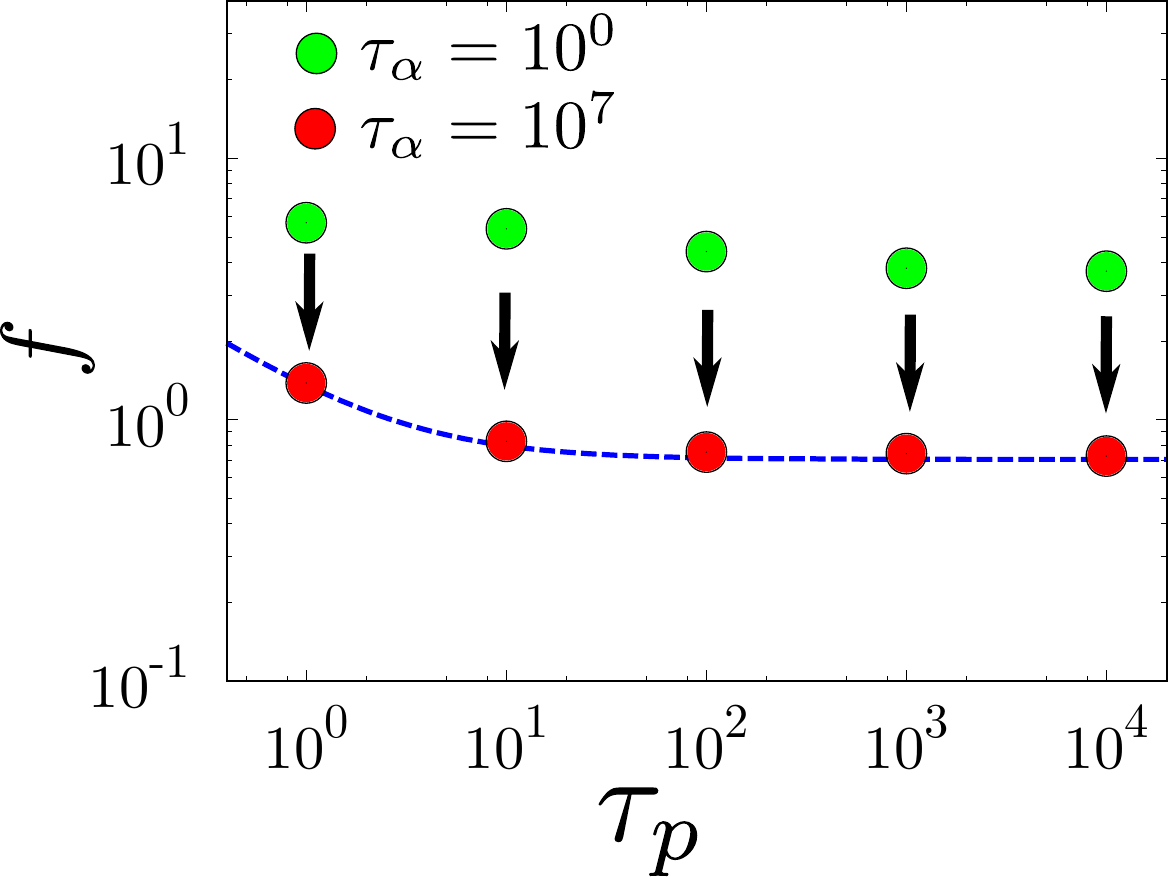}
\caption{(color online) Schematic of the quench protocol for the aging simulations. The green points (where $\tau_\alpha=1$) indicate at which $f$ the system is equilibrated initially; the red points (where the extrapolated relaxation time is $\tau_\alpha=10^7$) give the value of the active forcing $f$ that the system is quenched to for each $\tau_p$. The blue dotted line is a guide to eye.}
\label{fig:protocol}
\end{figure}

For each persistence time we first prepare the system at a value of the active forcing $f$ where the system behaves as an active liquid and the corresponding $\alpha$-relaxation time is $\tau_\alpha=1$ (using the definition $Q(\tau_{\alpha})=1/e$; no $t_{\rm w}$-dependence appears here as we are concerned with the behaviour in the stationary state). Then we perform a sudden quench to the value of $f$ where the extrapolated $\alpha$-relaxation time is $\tau_\alpha=10^7$ (see Fig.~\ref{fig:protocol} for a schematic of the protocol). We follow the same protocol for all values of $\tau_p$ except $\tau_p=\infty$. For the latter we quench from an analogous initial point (prepared at $\tau_p\to\infty$ and $f=2.95$ such that the corresponding relaxation time is again $\tau_\alpha=1$) to different values of the active forcing $f$ below $f_c$, where $f_c\approx 1.6$ is the jamming threshold for the model at $\tau_p\to\infty$. For intermediate persistence time $\tau_p=10^4$, the above protocol means in concrete terms that we prepare the initial conditions at $f=3.70$ ($\tau_\alpha=1$) and then quench to $f=0.72$ (extrapolated relaxation time $\tau_\alpha=10^7$).

\subsection{Small Persistence Time}

\begin{figure}
\centering
\includegraphics[height = 0.62\linewidth]{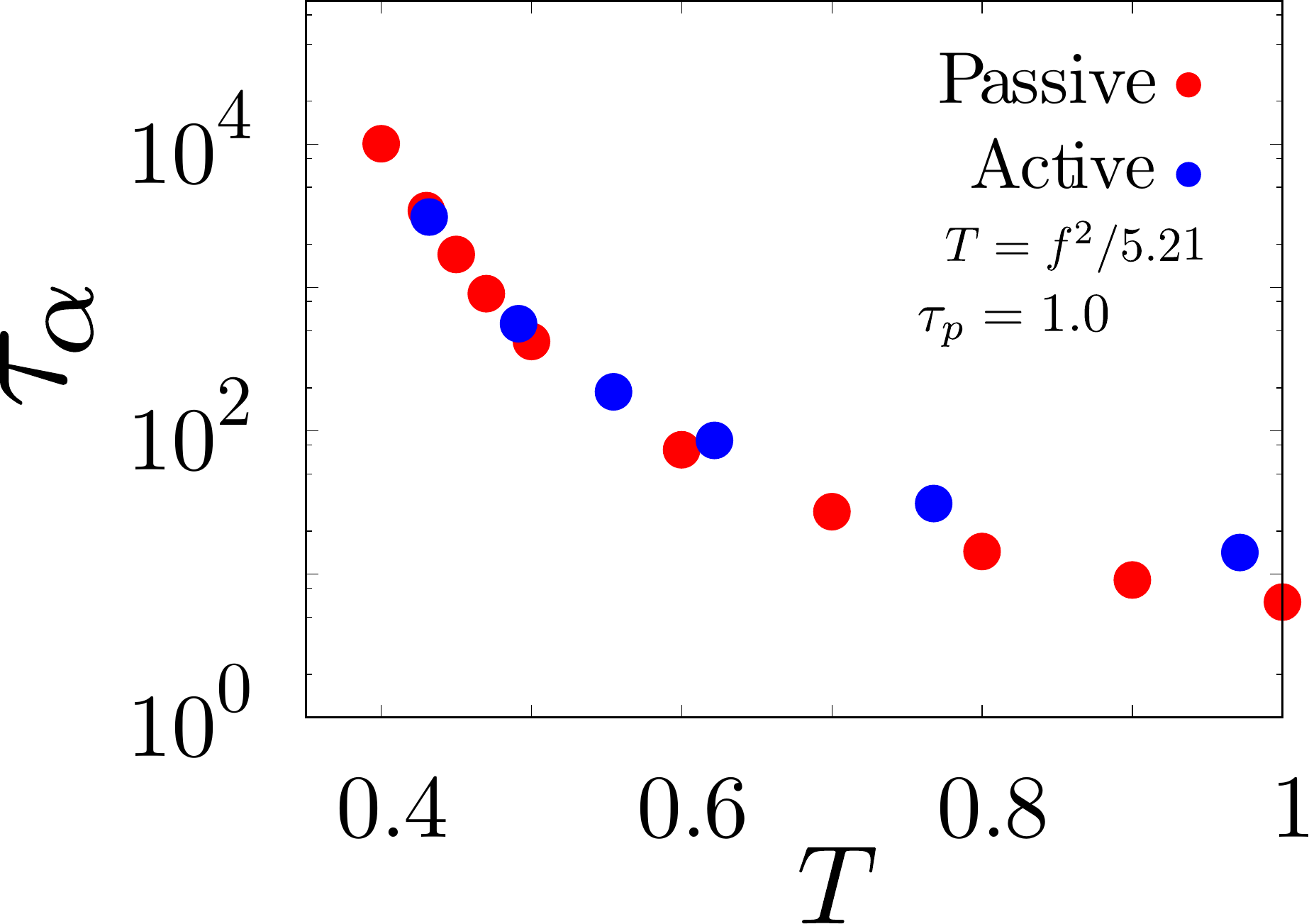}
\caption{(color online) Comparison of $\alpha$-relaxation time scale for a passive glass and the active glass with small persistence time $\tau_p=1.0$; for the active case we show on the $x$-axis the effective temperature calculated using the form $T_{\rm eff}=c f^2 \tau_p$ with $c \approx 0.2$.}
\label{fig:timescale}
\end{figure}

In the small persistence time limit we first identify the mapping between active forcing $f$ and temperature $T$. We carry out simulations for the active model glass for small persistence time, $\tau_p=1$, measuring $\tau_\alpha$ for different values of active forcing $f$. We also perform simulations of the corresponding passive two-dimensional Kob Andersen glass at different temperatures. A simple scaling from $f$ to $T$ of the form $T_{\rm eff}=c f^2 \tau_p$ with a constant factor ($c \approx 0.2$) collapses the $\tau_\alpha$-variation in the two cases (see Fig.~\ref{fig:timescale}).
Hence the equivalent post-quench point that corresponds to the active simulation ($f=1.38$ for $\tau_p=1$) becomes $T=0.37$ for the passive case.

\begin{figure}
\centering
\includegraphics[height = 0.62\linewidth]{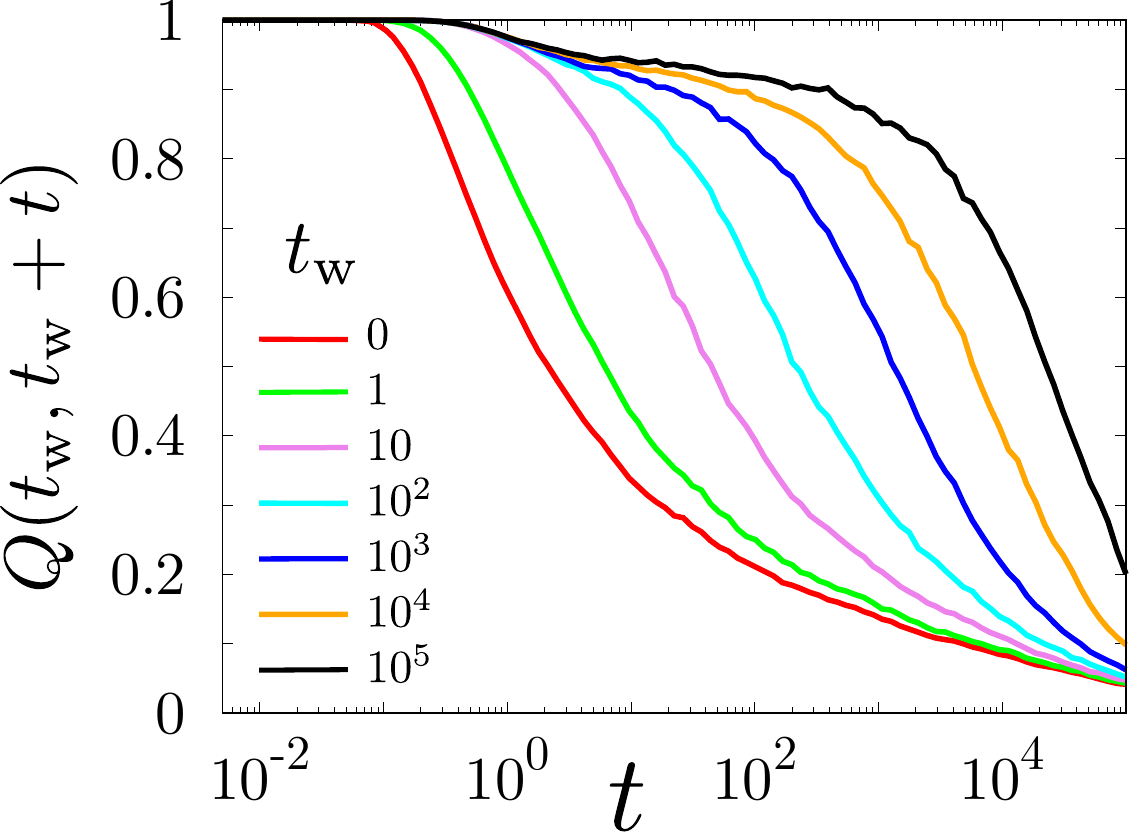}
\caption{(color online) Two-time overlap function $Q(t_{\rm w},t_{\rm w}+t)$ for a passive system quenched to $T=0.37$, plotted as a function of time $t$ for different waiting times $t_{\rm w}$ as shown.}
\label{fig:qtpassive}
\end{figure}

\begin{figure}
\centering
\includegraphics[height = 1.2\linewidth]{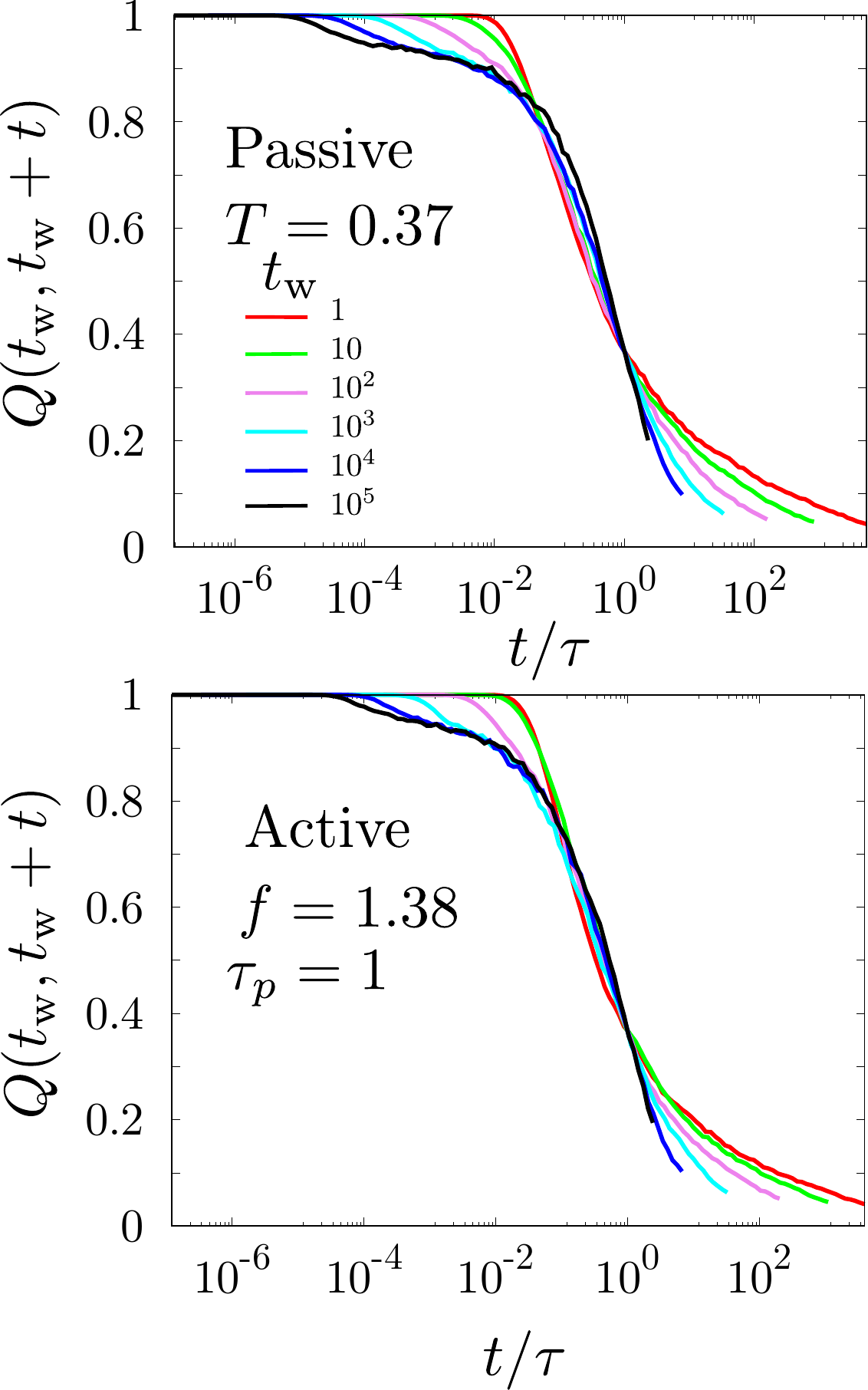}
\caption{(color online) Two-time overlap function $Q(t_{\rm w},t_{\rm w}+t)$ for the passive system (quenched to $T=0.37$) and the active system ($\tau_p=1$, quenched to $f=1.38$) plotted against rescaled time $t/\tau$ with $\tau=t^{0.7}_{\rm w}$. The close similarity between the figures demonstrates that the active system at small persistence time exhibits effectively thermal aging.}
\label{fig:qtsmalltaucom}
\end{figure}

Having identified the mapping between passive and short $\tau_p$-active systems, we verify that this mapping also captures the aging of the active system. In Fig.~\ref{fig:qtpassive} we show the two-time overlap function $Q(t_{\rm w},t_{\rm w}+t)$ for different waiting times $t_{\rm w}$ after a quench to $T=0.37$, which is almost identical to Fig.{\red{1}} for the active case in the main text. The aging collapse of the $Q(t_{\rm w},t_{\rm w}+t)$ data (see Fig.~\ref{fig:qtsmalltaucom}) is also very similar for the passive (quench to $T=0.37$) and active  ($\tau_p=1$, quench to $f=1.38$) systems; here the time difference $t$ has been scaled by $t^{0.7}_{\rm w}$ in both systems.

\subsection{Infinite Persistence Time}

\begin{figure}
\centering
\includegraphics[height = 1.2\linewidth]{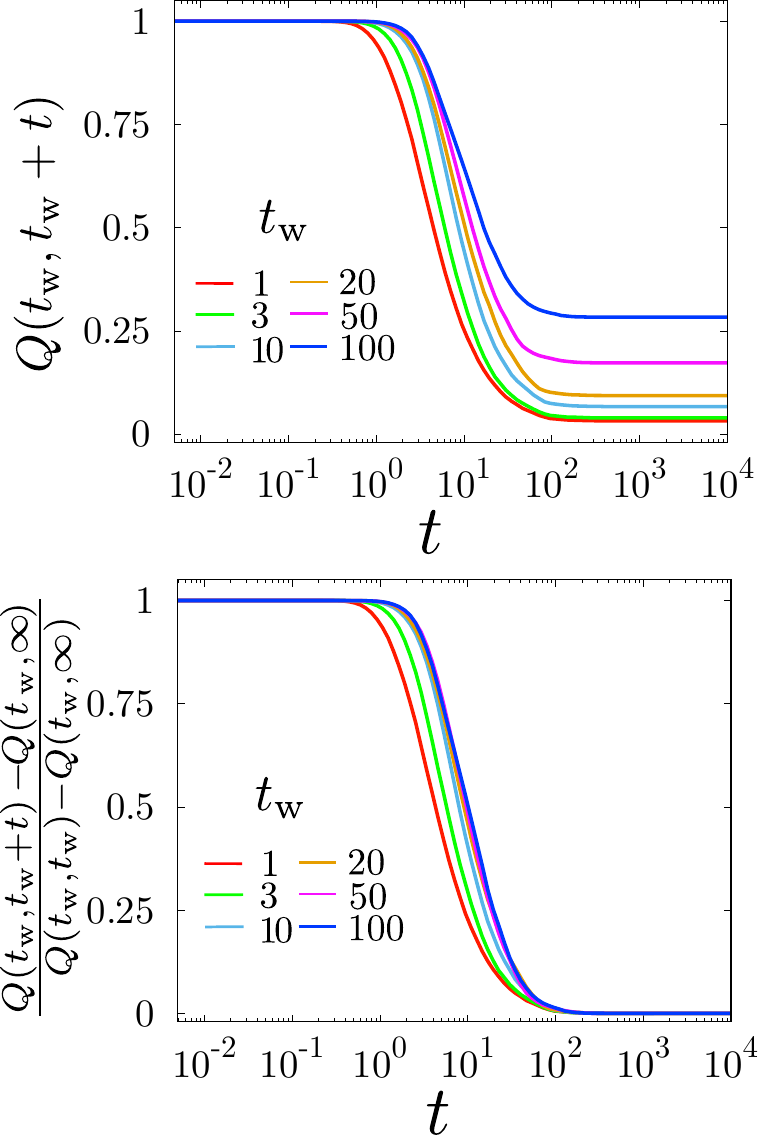}
\caption{(color online) (Top) Two-time overlap function for a quench to $f=1.2$ at $\tau_p=\infty$. (Bottom) Scaled two-time overlap function, which by definition decays from $1$ to $0$.}
\label{fig:qttauinf}
\end{figure}

To extract the aging or waiting time dependence of the overlap function for infinite persistence time we perform the following scaling. $Q(t_{\rm w},t_{\rm w}+t)$ for $\tau_p\to \infty$ decays to a plateau from which it does not relax further (see Fig.~\ref{fig:qttauinf}). Therefore we calculate
\begin{equation}
    C(t,t_{\rm w})=\frac{Q(t_{\rm w},t_{\rm w}+t)-Q(t_{\rm w},\infty)}{Q(t_{\rm w},t_{\rm w})-Q(t_{\rm w},\infty)}
\end{equation} where $Q(t_{\rm w},\infty)$ is the plateau height of the overlap function for a given $t_{\rm w}$. The function $C(t,t_{\rm w})$ thus starts out at unity and decays to zero (see Fig.~\ref{fig:qttauinf}). We then extract a relaxation timescale from this function using the conventional definition $C(\tau,t_{\rm w})=1/e$.

\begin{figure}
\centering
\includegraphics[height = 0.65\linewidth]{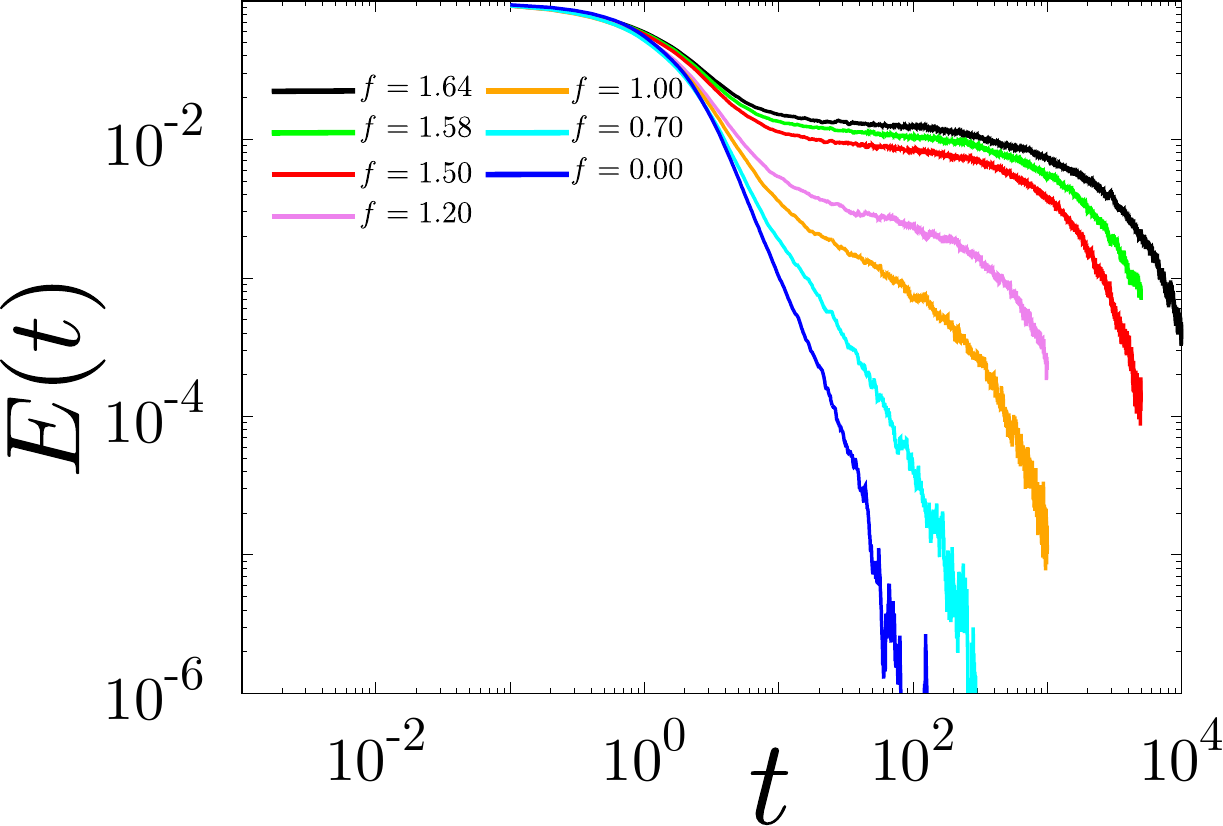}
\caption{(color online) Kinetic energy decay for infinite persistence time, for quenches to different values of the active forcing $f$. For moderate $f$ the decay of the kinetic energy follows a power law while as $f$ approaches $f_c$ a plateau develops before the final decay.}
\label{fig:energy}
\end{figure}

The decay in the kinetic energy during the aging process follows a power law for the athermal aging reported in Ref.~\cite{chacko19}. We observe similar behaviour for moderate $f$ but as $f$ increases towards $f_c$ the decay exponent becomes smaller and the power-law decay crosses over to a plateau (see Fig.~\ref{fig:energy}) that is followed by a final decay at later times. The kinetic energy plateau suggests that the physics here has aspects similar to creep-like dynamics, where a system held below its yield stress can flow transiently for some time before arresting~\cite{barrat18,liu18}.

\subsection{Intermediate Persistence Time}

\begin{figure}
\centering
\includegraphics[height = 0.65\linewidth]{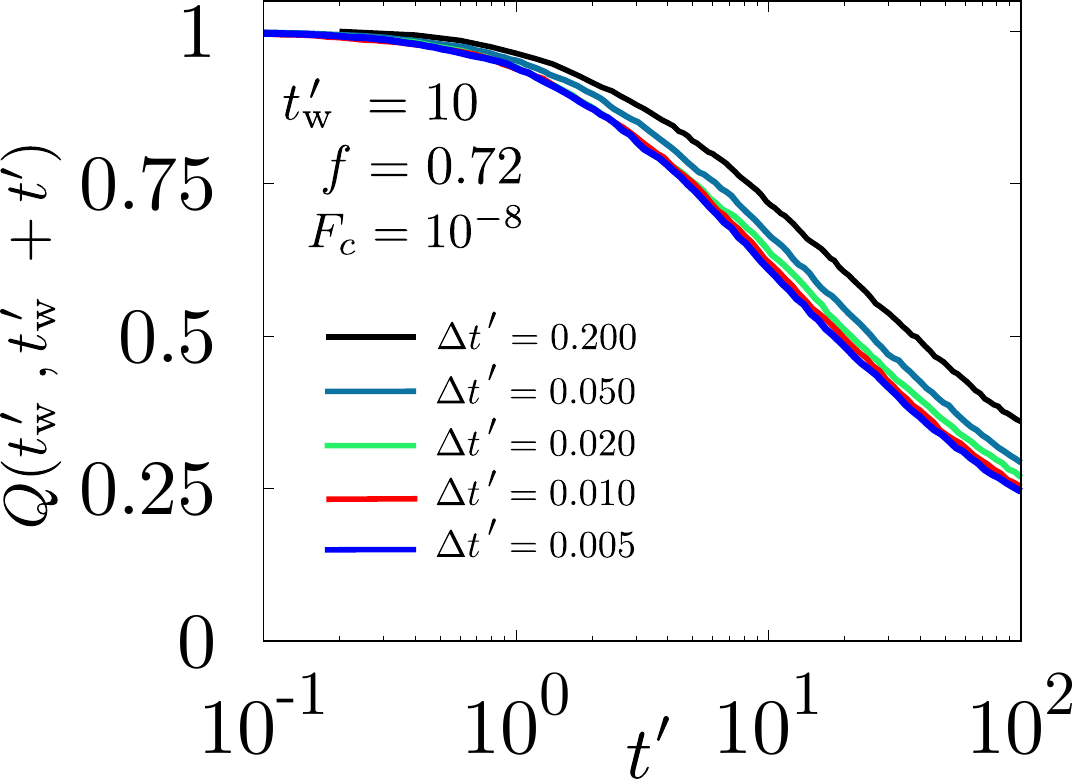}
\caption{(color online) Convergence of activity driven dynamics with respect to timestep $\Delta t^{\prime}$, showing two-time overlap function for $t_{\rm w}'=10$ after a quench to $f=0.72$ for different values of $\Delta t^{\prime}$ at fixed force threshold $F_c=10^{-8}$.}
\label{fig:ADAconvergence1}
\end{figure}

\begin{figure}
\centering
\includegraphics[height = 0.65\linewidth]{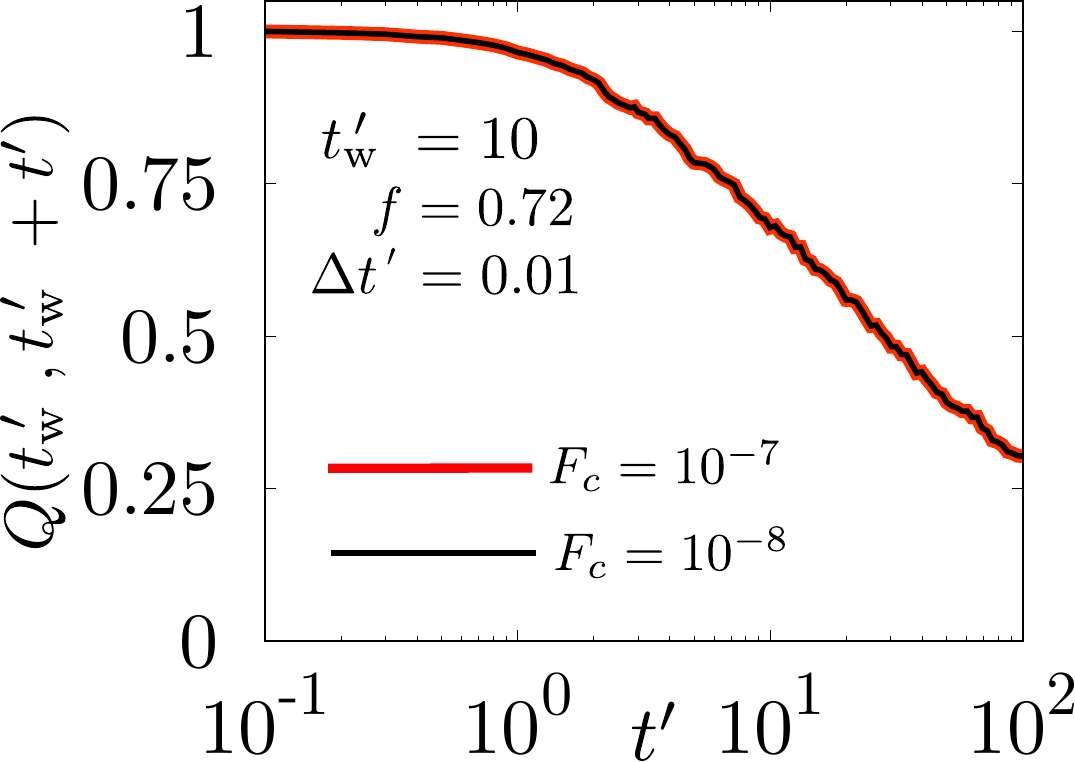}
\caption{(color online) Convergence of activity driven dynamics with respect to force threshold, showing analogous results to Fig.~\ref{fig:ADAconvergence1} for $\Delta t^{\prime}=0.01$ and two different values of the force threshold $F_c$ used during the adaptive minimization.}
\label{fig:ADAconvergence2}
\end{figure}

We first check convergence of the algorithm for activity driven dynamics as described in the main text. To reach force-free states numerically, we use an adaptive minimization protocol that stops when root mean square force lies below a threshold $F_c$. The dynamics is then driven by making small changes in the orientation  $\theta_i$ of the propulsion force for each particle, for fixed small steps of rescaled time $\Delta t^{\prime}$. We first vary the time step $\Delta t^{\prime}$, for a fixed force threshold $F_c=10^{-8}$, and then later explore the dependence on $F_c$.
We find that for $\Delta t^{\prime}=0.01$ or below the two-time overlap function $Q(t^{\prime}_{\rm w}, t^{\prime}_{\rm w}+t^{\prime})$ becomes independent (see Fig.~\ref{fig:ADAconvergence1}) of the time step $\Delta t^{\prime}$ used. Keeping then $\Delta t^{\prime}=0.01$, we change the inter-particle force threshold $F_c$ and observe that the results for $Q(t^{\prime}_{\rm w}, t^{\prime}_{\rm w}+t^{\prime})$ become $F_c$-independent for $F_c\leq 10^{-7}$ (see Fig.~\ref{fig:ADAconvergence2}). We therefore use values well within this converged region, $F_c=10^{-8}$ and $\Delta t^{\prime}=10^{-2}$, for all the activity driven dynamics runs presented in the main text.

\begin{figure}
\centering
\includegraphics[height = 0.65\linewidth]{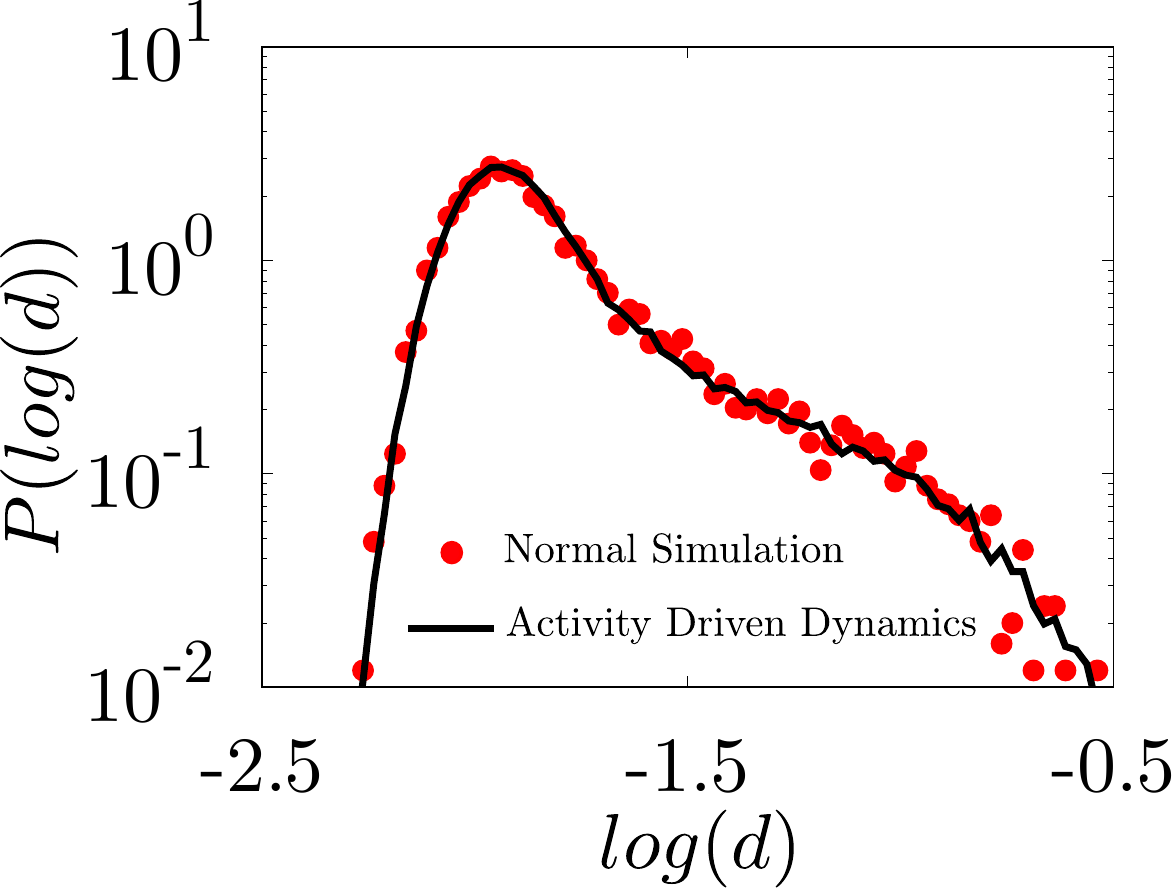}
\caption{(color online) Comparison of probability distributions of logarithmic displacements between activity driven dynamics and normal simulations with $\tau_p=10^4$, both for steady state dynamics at $f=1.0$.}
\label{fig:pdfdis}
\end{figure}

To cross check that the new algorithm of activity driven dynamics that faithfully reproduces the dynamical behaviour of the original dynamics at large $\tau_p$, we further analyse the intermittent time series data of displacements (see Fig.~{\red{5}} in the main text) and calculate the probability distribution of displacements (measured as root mean squared displacements calculated over a time window $\delta t^{\prime}=0.01$ for ADA and the corresponding $\delta t=100$ for $\tau_p=10^4$ in a standard simulation). Fig.~\ref{fig:pdfdis} demonstrates that these distributions agree, thus providing quantitative confirmation of the visual similarity of the displacement time series in Fig.~{\red{5}} in the main text.

\begin{figure}
\centering
\includegraphics[height = 0.65\linewidth]{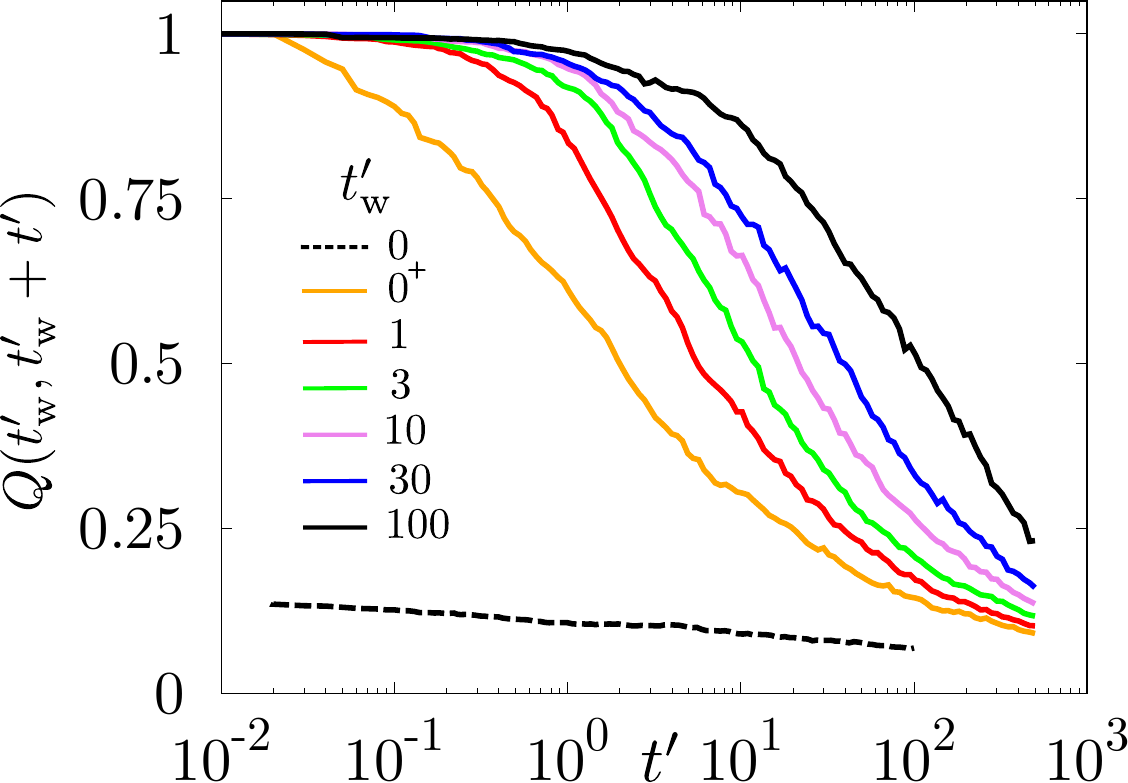}
\caption{(color online) Two-time overlap function $Q(t^{\prime}_{\rm w}, t^{\prime}_{\rm w}+t^{\prime})$ for activity driven aging dynamics for different $t^{\prime}_{\rm w}$.}
\label{fig:addall}
\end{figure}

Finally for the activity driven dynamics we plot in Fig.~\ref{fig:addall} the overlap function $Q(t^{\prime}_{\rm w}, t^{\prime}_{\rm w}+t^{\prime})$ for different $t_{\rm w}'$ including $t_{\rm w}'=0$ and $t_{\rm w}'=0^+$. The results illustrate the difference between strictly zero age (where the reference configuration at $t_{\rm w}'=0$ is {\em not} force-free) and nonzero but small scaled age $t_{\rm w}'$. In the latter case the reference configuration {\em is} force-free and further activity-driven evolution, i.e.\ increasing $t'$ decreases the overlap $Q$ gradually from unity. In the $t_{\rm w}'=0$ case, on the other hand, time evolution of any duration $t'>0$ leads to a force-free configuration that is distinct from the initial one, giving a discontinuous drop in $Q$. 

